\documentclass[twocolumn,iop,apj,numberedappendix,appendixfloats,twocolappendix]{openjournal}
\pdfoutput=1 
\usepackage{anyfontsize}
\usepackage[T1]{fontenc}
\usepackage[utf8]{inputenc}
\usepackage[english]{babel}
\usepackage{newtx}
\usepackage{amsmath,amstext}
\usepackage{microtype}
\usepackage[dvipsnames]{xcolor}
\usepackage[pdftex,breaklinks,colorlinks,allcolors=Blue,pdfusetitle]{hyperref}
\usepackage[figure,figure*]{hypcap}
\usepackage{orcidlink}

\input{hyperlink-year-only-natbib-patch}
\allowdisplaybreaks
\graphicspath{{figures/}}
\renewcommand*{\baselinestretch}{1.1}

\defcitealias{Kravtsov:2013}{K13}
\defcitealias{Hearin:2019}{H19}

\newcommand*{\kms}{\text{km}\,\text{s}\ensuremath{^{-1}}}
\newcommand*{\msun}{\ensuremath{M_{\odot}}}
\newcommand*{\mstar}{\ensuremath{M_{\star}}}
\newcommand*{\perh}{\ensuremath{h^{-1}}}
\newcommand*{\LCDM}{\ensuremath{\Lambda}\text{CDM}}

\newcommand*{\mpeak}{\ensuremath{M_\text{peak}}}
\newcommand*{\mvir}{\ensuremath{M_\text{vir}}}
\newcommand*{\rvir}{\ensuremath{r_\text{vir}}} 
\newcommand*{\ampeak}{\ensuremath{a_{\mpeak}}}
\newcommand*{\rmpeak}{\ensuremath{r_{\mpeak}}}

\newcommand*{\rhalf}{\ensuremath{r_{1/2}}}
\newcommand*{\logmstar}{\ensuremath{\log{\left(\mstar/\msun\right)}}}

\newcommand*{\code}[1]{\ensuremath{\mathtt{#1}}}
\newcommand*{\https}[1]{\href{https://#1}{\nolinkurl{#1}}}
\newcommand*{\http}[1]{\href{http://#1}{\nolinkurl{#1}}}

\shorttitle{Galaxy--halo Size Relations Impact Galaxy Clustering}
\shortauthors{Hill and Mao}

\makeatletter
\def\frontmatter@title@below{\vspace*{-0.5em}}
\hypersetup{pdfauthor={\@rectohead}}
\makeatother

\begin{document}

\renewcommand*{\sectionautorefname}{Section} 
\renewcommand*{\subsectionautorefname}{Section} 
\renewcommand*{\subsubsectionautorefname}{Section} 

\title{The Impact of Galaxy--halo Size Relations on Galaxy Clustering Signals}

\author{Joshua~B.~Hill\,\orcidlink{0000-0001-9764-7385}}
\author{Yao-Yuan~Mao\,\orcidlink{0000-0002-1200-0820}}
\affiliation{Department of Physics and Astronomy, University of Utah, Salt Lake City, UT 84112, USA}

\begin{abstract}
Galaxies come in different sizes and morphologies, and these differences are thought to correlate with properties of their underlying dark matter halos. However, identifying the specific halo property that controls the galaxy size is a challenging task, especially because most halo properties depend on one another. In this work, we demonstrate this challenge by studying how the galaxy--halo size relations impact the galaxy clustering signals. We investigate the reason that a simple linear relation model, which prescribes that the galaxy size is linearly proportional to the dark matter halo's virial radius, can still produce clustering signals that match the observational data reasonably well. We find that this simple linear relation model for galaxy sizes, when combined with the subhalo abundance matching technique, introduces an implicit dependence on the halo formation history. As a result, the effect of halo assembly bias enters the resulting galaxy clustering, especially at lower stellar masses, producing a clustering signal that resembles the observed one.  At higher stellar masses, the effect of halo assembly bias weakens and is partially canceled out by the effect of halo bias, and the clustering of large and small galaxies becomes more similar. This combined effect implies that small and large galaxies not only occupy halos of different masses, but they must also occupy halos of different assembly histories. Our study highlights the challenge of identifying a particular halo property that controls galaxy sizes through constraints from galaxy clustering alone. 
\end{abstract}

\keywords{%
\href{http://astrothesaurus.org/uat/356}{Dark matter distribution (356)},
\href{http://astrothesaurus.org/uat/617}{Galaxy radii (617)},
\href{http://astrothesaurus.org/uat/902}{Large-scale structure of the universe (902)},
\href{http://astrothesaurus.org/uat/1083}{N-body simulations (1083)},
\href{http://astrothesaurus.org/uat/1880}{Galaxy dark matter halos (1880)}%
}

\section{Introduction}

In the Lambda--Cold Dark Matter (\LCDM{}) cosmological model, structures form hierarchically, driven by the gravitational collapse of dark matter overdensities, forming dark matter halos. These halos arise from the small density fluctuations in the early universe, and these fluctuations grow as the universe expands. As such, dark matter halos are biased tracers of the dark matter density field \citep{2005MNRAS.363L..66G,0801.4826}. Over time, smaller halos merge to form larger ones, creating a hierarchical structure. Dark matter halos serve as the gravitational environment upon which galaxies form and evolve. Galaxies interact with their surrounding dark matter halos through gravitational forces, influencing their evolution and dynamics. Thus the spatial distribution of galaxies is a prime observable to test theoretical models of dark matter halos and galaxy formation.

Many studies have sought to link halo properties with galaxy properties (see \citealt{1804.03097} for a review). These galaxy--halo connection models can be established through hydrodynamical simulations, or via empirical models that directly map simulated halo properties onto observed galaxy quantities. 
One of the main achievements in this field is the establishment of the stellar-to-halo mass relation (SHMR) \citep[e.g.,][]{0903.4682,1806.07893}. Beyond mass, models of the galaxy--halo connection also explore relationships between other galaxy and halo properties, such as galaxy color and size \citep[e.g.,][]{2013MNRAS.435.1313H,2018MNRAS.473.2714S}. 

Galaxy sizes, in particular, have been of recent interest both theoretically \citep{Almeida:2020,Rodriguez:2021,Klein:2024} and observationally \citep{Du:2024,Ito:2024,Buitrago:2024,2106.14924}. One major question on this topic is what halo properties have the strongest influence on galaxy size. 
For instance, \citet[hereafter \citetalias{Kravtsov:2013}]{Kravtsov:2013} finds that by combining a linear scaling relation between galaxy and halo radii and the subhalo abundance matching technique, a technique commonly used in building a SHMR model \citep{Kravtsov:2004,1207.2160}, one can reproduce the observed galaxy stellar mass--size relation (see also \citealt{Huang:2017} for higher-redshift results and \citealt{2015MNRAS.454..322D} for scatter in this relation). The other class of galaxy size models, motivated by the physics of disk formation, is based on halo spin or angular momentum \citep{1980MNRAS.193..189F.1998MNRAS.295..319M}. This class of models is often used in semi-analytical galaxy formation models \citep[e.g.,][]{2008MNRAS.391..481S,2012NewA...17..175B}.

There have also been many studies that attempted to constrain the galaxy size models by observations beyond the galaxy stellar mass--size relation alone, most commonly with galaxy clustering. For example, \citet[hereafter \citetalias{Hearin:2019}]{Hearin:2019} uses the \citetalias{Kravtsov:2013} model to predict galaxy clustering of galaxies of different sizes, comparing the results with observations from the Sloan Digital Sky Survey (SDSS; \citealt{1608.02013}). After employing both the subhalo abundance matching technique and the \citetalias{Kravtsov:2013} model, they categorized the mock galaxies as large or small based on whether their sizes were above or below the median at a fixed stellar mass and compared their two-point correlation functions. They found that the small galaxies cluster more than the large galaxies at low stellar masses, but both samples cluster similarly at high stellar masses. This behavior matches reasonably well with the clustering observed in SDSS galaxies when split into large and small samples as well. 

However, more recent studies by \citet{2019MNRAS.488.4801J} and \citet{2101.05280} show that, in order to match the observed two-point correlation functions to a greater degree, further dependence on halo formation histories needs to be introduced to these galaxy size models. Specifically, \citet{2019MNRAS.488.4801J} introduced a halo concentration dependence to the proportionality constant of the linear relation between galaxy and halo radii, and \citet{2101.05280} introduced a new model where galaxy sizes are proportional to the halo growth rate. 
In both cases, the updated models produce two-point correlation functions that better match to SDSS data, compared to the earlier size models. 

The idea that galaxy sizes would depend on properties beyond just the halo radius is not really surprising; after all, galaxies are strongly influenced by the formation history of their host halos. What might be surprising is that the linear relation model (e.g., the \citetalias{Kravtsov:2013} model), while not matching the data perfectly, still works reasonably well. 

In this work, we aim to provide an explanation of why the linear relation model can produce a reasonable clustering signal, as found by \citetalias{Hearin:2019}. We argue that, although the linear relation model appears to connect the galaxy radius directly to the halo radius when combined with the subhalo abundance matching technique for generating stellar masses, the galaxy size model implicitly makes use of the halo formation history information. As a result, the effect of halo assembly bias \citep[cf.][]{2005MNRAS.363L..66G,Wechsler:2006,2018MNRAS.474.5143M} still enters the galaxy--halo size relation and impacts the resulting galaxy clustering. 

This paper is structured as follows, we introduce the data and methods, which follow \citetalias{Hearin:2019} closely, in \autoref{sec:methods}. We present our main results and theoretical explanation of clustering in \autoref{sec:results}. Finally, we discuss the implication of our results by showcasing a comparison between two similar models in \autoref{sec:discussion}.

\section{Methods}
\label{sec:methods}

\subsection{Simulations}
\label{sec:sims}
This work uses the Very Small Multi-Dark Planck Simulation (VSMDPL; \citealt{2013AN....334..691R}), which is a cosmological, gravity-only, $N$-body simulation. The VSDMPL simulation has a periodic boundary of 160\,Mpc\,\perh, with $3840^3$ particles. The mass resolution for each particle is $6.18 \times 10^6$\,\msun\,\perh, while the physical force resolution is 5\,kpc\,\perh. The VSDMPL simulation follows the \citet{PlanckCollaboration:2013} \LCDM{} cosmology, with the following parameters: $\Omega_m = 0.307115$, $\Omega_\Lambda = 1 - \Omega_m$, $\Omega_b = 0.048206$, Hubble parameter today is $H_0 = 100 \, h$\,\kms\,Mpc$^{-1}$ with $h = 0.6777$, scalar spectral index $n_s = 0.96$, and the amplitude of mass density fluctuation $\sigma_8 = 0.8228$. 
We use the halo catalogs generated by \code{Rockstar} and \code{ConsistentTrees} \citep{2013ApJ...762..109B,1110.4370}. 
Note that we have converted the $h$-scaled mass and distance units (i.e., $h=1$ units) from the simulation to $h = 0.6777$ units, unless otherwise specified. 

\subsection{Model}
\label{sec:model}
The model setup of this work closely follows the procedure of \citet{Hearin:2019}, with the exception that we do not include orphan galaxies (galaxies whose dark matter halos have already been completely disrupted).  The model has two components to assign galaxy stellar mass and sizes. First, a standard abundance matching procedure is used to assign a galaxy stellar mass to each (sub)halo. Second, the half-light radius ($\rhalf$) for each galaxy is calculated by the \citetalias{Kravtsov:2013} model. These two model components are applied independently; that is, the model does not include an explicit relationship between stellar masses and galaxy sizes. The correlation between the modeled stellar masses and galaxy sizes stem from the underlying correlations among the halo properties used in the model.
We describe these two components in detail below.

\subsubsection{Galaxy Stellar Masses}
\label{sec:SHAM}
The galaxy stellar masses are assigned to halos based on the subhalo abundance matching technique, which maps galaxy stellar mass, $\mstar$, to halos of a given mass. The technique is based on the simple idea that the most luminous or massive galaxy will live in the most massive (sub)halo, and the second most massive galaxy lives in the second most massive (sub)halo, and so on, with volume correction and some element of scatter \citep{2010ApJ...717..379B,1207.2160}. The subhalo abundance matching technique can be applied to halo properties other than mass. In fact, studies have shown that the peak maximum circular velocity along the halo's main branch works the best to reproduce galaxy clustering \citep[e.g.,][]{1207.2160,2016MNRAS.460.3100C,2017ApJ...834...37L}. 

Here, we follow the choices made by \citet{Hearin:2019} and use halo \mpeak, the highest \mvir{} a halo ever reaches on its main branch, as the matching property. 
We assume a $0.2$ dex of scatter in the subhalo abundance matching procedure.  We match halos with a galaxy of a given stellar mass using the stellar mass functions given in \citet[for $z = 0$ and 0.5]{Moustakas:2013} and \citet[for $z = 1$, 2, and 3]{2011MNRAS.413.2845M}. The subhalo abundance matching step is implemented through a publicly available abundance matching code \citep{2010ApJ...717..379B,2022ascl.soft12016M}. 

\subsubsection{Galaxy Sizes}
\label{sec:galmodel}
To assign galaxy sizes, we apply the \citetalias{Kravtsov:2013} model, which prescribes a relationship between the half-light radius (\rhalf) of a galaxy and the viral radius of a halo following a simple linear relationship, 
$\rhalf{} = 0.01 \rvir$. \citetalias{Hearin:2019} chooses to extend this model to be 
\begin{equation}
\rhalf{} = 0.01 \rmpeak,
\label{eq:rhalf}
\end{equation}
where $\rmpeak{} = \rvir(\ampeak)$ is the viral radius of a halo at scale factor when the halo reaches its peak mass, or \ampeak. 
The relationship between \mpeak{} and \rmpeak{} is,
\begin{equation}
\mpeak{} = \frac{4\pi}{3} \rmpeak^3 \Delta_\text{vir}\left(\ampeak\right)\rho_\text{crit}\left(\ampeak\right),
\label{eq:mvir}
\end{equation}
where the spherical overdensity $\Delta_\text{vir}$ has a dependence on the scale factor,
\begin{equation}
\Delta_\text{vir}(a) = 18\pi^2 + 82\left(\Omega_m\left(a\right)-1\right) - 39\left(\Omega_m\left(a\right)-1\right)^2
\label{eq:Delta}
\end{equation}
Where $\Omega_m\left(a\right)$ and $\rho_\text{crit}\left(a\right)$ are determined by the Hubble constant when each halo reaches its peak mass $H(\ampeak)$.  

\begin{figure}[tb!]
\centering
\includegraphics[width=\linewidth]{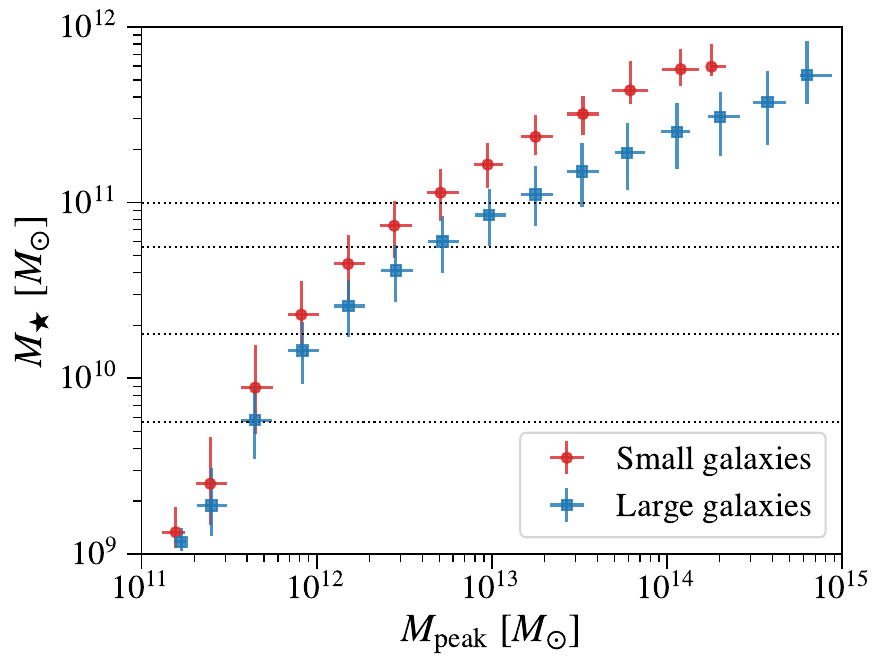}
\caption{Median stellar-to-halo mass relation for model-predicted small (\textit{red-filled circles}) and large (\textit{blue-filled squares}) galaxies. The designations of small and large galaxies are described in \autoref{sec:galclustering}. Each point shows the median stellar mass (\mstar) and peak halo mass (\mpeak{}) in each \mpeak{} bin, with the error bars showing $1\sigma$ of \mstar{} and \mpeak{} distributions in that bin. The four horizontal dotted lines denote the four \mstar{} thresholds for the clustering analysis: $\logmstar \ge 9.75$, 10.25, 10.75, and 11. 
\label{fig:SMHR}}
\end{figure}

\subsection{Galaxy Samples and Clustering}
\label{sec:galclustering}

The modeled galaxies are split into small and large subsamples based on their model-predicted half-light radii. We follow the same procedure in \citetalias{Hearin:2019}. We first find the median of the galaxy half-light radii as a function of stellar mass by running a sliding window in small bins of stellar mass. Then, any galaxy whose radius lies above the median radius at its stellar mass will be designated as a large galaxy. In contrast, a galaxy whose radius lies below will be designated as a small galaxy. 
\autoref{fig:SMHR} shows the SHMR for the small and large galaxy subsamples; however, note that the designations of small and large galaxies are based on the modeled radii, which is connected to the halo virial radius through \autoref{eq:rhalf} and to halo mass through \autoref{eq:mvir}.

At low stellar masses, where the SHMR slope is steeper, a narrow stellar mass bin corresponds to a narrow range of halo masses. As such, the halo masses of small and large galaxies cannot differ too much, since the halo mass range is narrow to begin with. However, as the SHMR slope becomes significantly shallower at higher stellar masses, a narrow stellar mass bin now corresponds to a wider range of halo masses. In this regime, when we split the galaxy sample into large and small subsamples, the large galaxies will tend to have higher halo mass.

For most of our analyses, we also split the galaxies into four samples of different stellar masses, based on the following \mstar{} thresholds:  $\logmstar \ge 9.75$, 10.25, 10.75, and 11.  The horizontal dotted lines in \autoref{fig:SMHR} denote these four stellar mass thresholds. 

To analyze the relative clustering of large and small galaxies, we calculate the three-dimensional space two-point correlation function $\xi\left(r_p,\pi\right)$ using the estimator $\xi\left(r_p,\pi\right) = \frac{DD}{RR} - 1$ \citep{1982MNRAS.201..867H} for both large and small mock galaxies. In addition, we include peculiar velocities of the galaxies in the line of sight direction. Then to calculate the projected correlation function, $w_p\left(r_p\right)$, we integrate  $\xi\left(r_p,\pi\right)$:
\begin{equation}
w_p\left(r_p\right) = \int_{-\pi_\text{max}} ^{\pi_\text{max}} d\pi \, \xi\left(r_p,\pi\right), 
\label{eq:wp}
\end{equation}
where the line of sight projection, $\pi_\text{max}$, is 20\,Mpc\,\perh. 
We use the Halotools library \citep[\code{mock\_observables.wp};][]{1606.04106} to calculate $w_p\left(r_p\right)$. 

 Note that the two-point clustering statistics depend on the binning scheme in $r_p$. The binning scheme can reduce the effectiveness of the clustering statistics when we compare with the observational data. Using unbinned methods or high-order correlation functions can be more effective \citep{2408.16398}. In this work, we only focus on the relative clustering strength between the small and large galaxy samples. Therefore, the choice of binning scheme in $r_p$ does not significantly impact our analysis.

\section{Results}
\label{sec:results}

\begin{figure*}[tb!]
\centering
\includegraphics[width=\linewidth]{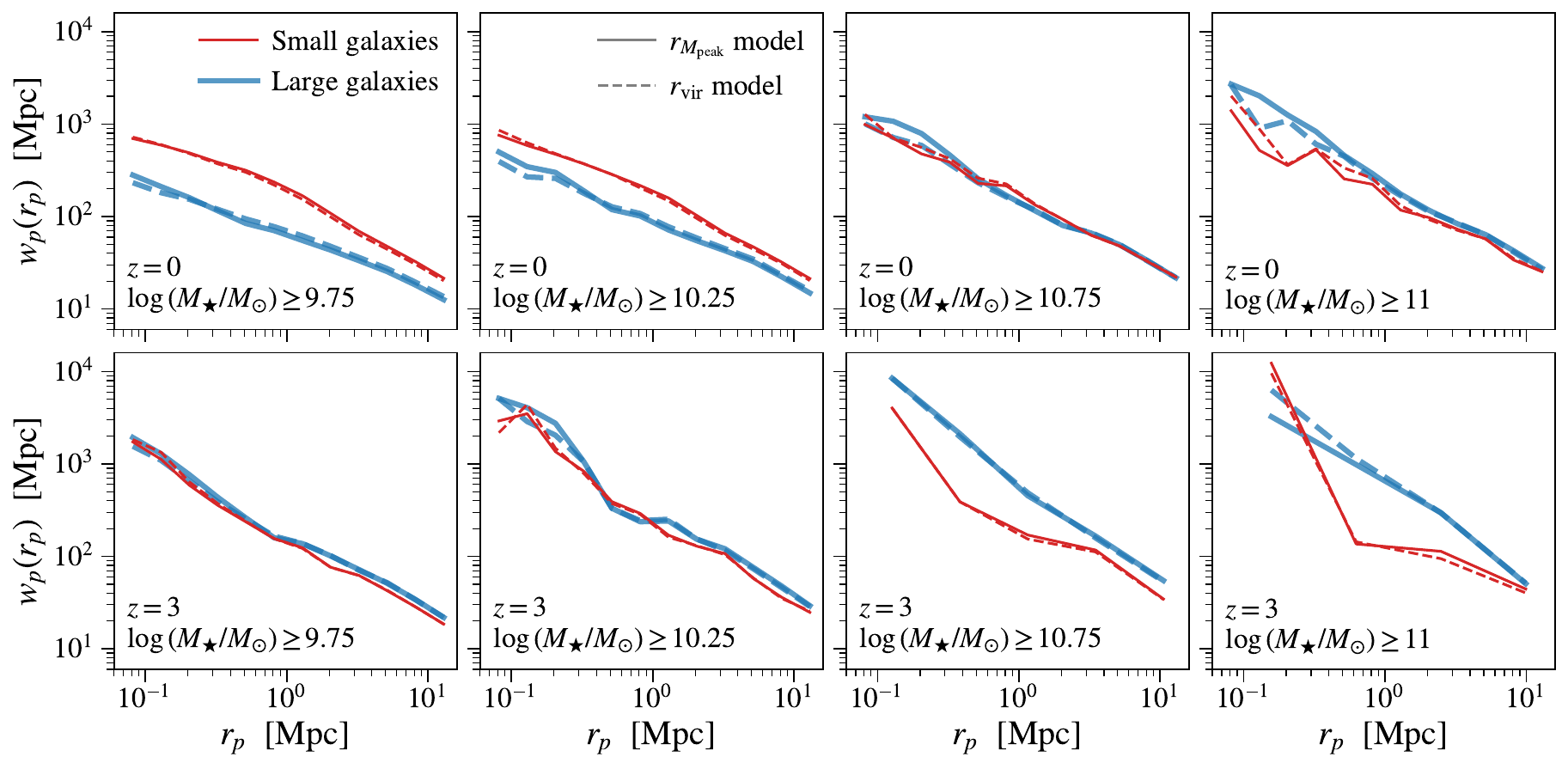}
\caption{Projected two-point (galaxy--galaxy) correlation function, $w_p(r_p)$, for various galaxy subsamples. Here $w_p$ and $r_p$ are both shown in comoving distances. 
The two rows show $w_p(r_p)$ at two redshifts: $z=0$ (\textit{upper}) and $z=3$ (\textit{lower}).
The four columns show $w_p(r_p)$ of four galaxy samples corresponding to different stellar mass thresholds (\textit{from left to right}): $\logmstar \ge 9.75$, 10.25, 10.75, and 11.
The color and thickness of the curves denote the two size-split subsamples: small (\textit{red thin curves}) and large (\textit{blue thick curves}) galaxies; the designations of small and large galaxies are described in \autoref{sec:galclustering}.
The line styles show the predictions from different size models: using \rmpeak{} (\textit{solid}) and \rvir{} (\textit{dashed}) in \autoref{eq:rhalf}.
\label{fig:clustering}
}
\end{figure*}

Using the methods discussed in \autoref{sec:methods}, we first reproduce the clustering signals for all the subsamples at $z=0$ in \citetalias{Hearin:2019}, which are shown in the upper panel of \autoref{fig:clustering}. 
At lower masses, the clustering signals for small galaxies are stronger (more clustered) than the larger ones. For higher-mass galaxies, on the other hand, large and small galaxies cluster similarly. 
These findings match those of \citetalias{Hearin:2019}, and qualitatively match the observed clustering signals from SDSS as shown in \citetalias{Hearin:2019}. 
Note that because we did not include orphan galaxies in our analysis, our clustering signals are slightly lower than those of \citetalias{Hearin:2019}. 

We then turn to the main result of this work: a detailed explanation of the $z=0$ clustering signals shown in \autoref{fig:clustering}. We find that the relative clustering of large and small galaxies stems mainly from the interplay between the relative halo bias and halo assembly bias.  We will show in \autoref{sec:assemblybias} that small galaxies exhibit a stronger halo assembly bias (dependence of clustering on halo assembly history at a fixed halo mass; see e.g., \citealt{2005MNRAS.363L..66G,Wechsler:2006,0801.4826}) , and in \autoref{sec:halobias} that the halo bias effect (dependence of clustering on halo mass; see e.g., \citealt{1999MNRAS.308..119S}) dominates for high-mass galaxies. The combination of both halo bias and assembly bias is discussed in \autoref{sec:botheffects}.

In \autoref{fig:clustering}, we also show the clustering signals for a slightly different galaxy size model that uses \rvir{} instead of \rmpeak{}. We will discuss its similarities and differences in \autoref{sec:rvir}. The lower panels of \autoref{fig:clustering} show the clustering signals at a higher redshift ($z=3$), which will be discussed further in \autoref{sec:highzpred}. 

\subsection{Assembly Bias Effect}
\label{sec:assemblybias}

\begin{figure*}[tb!]
\centering
\includegraphics[width=\linewidth]{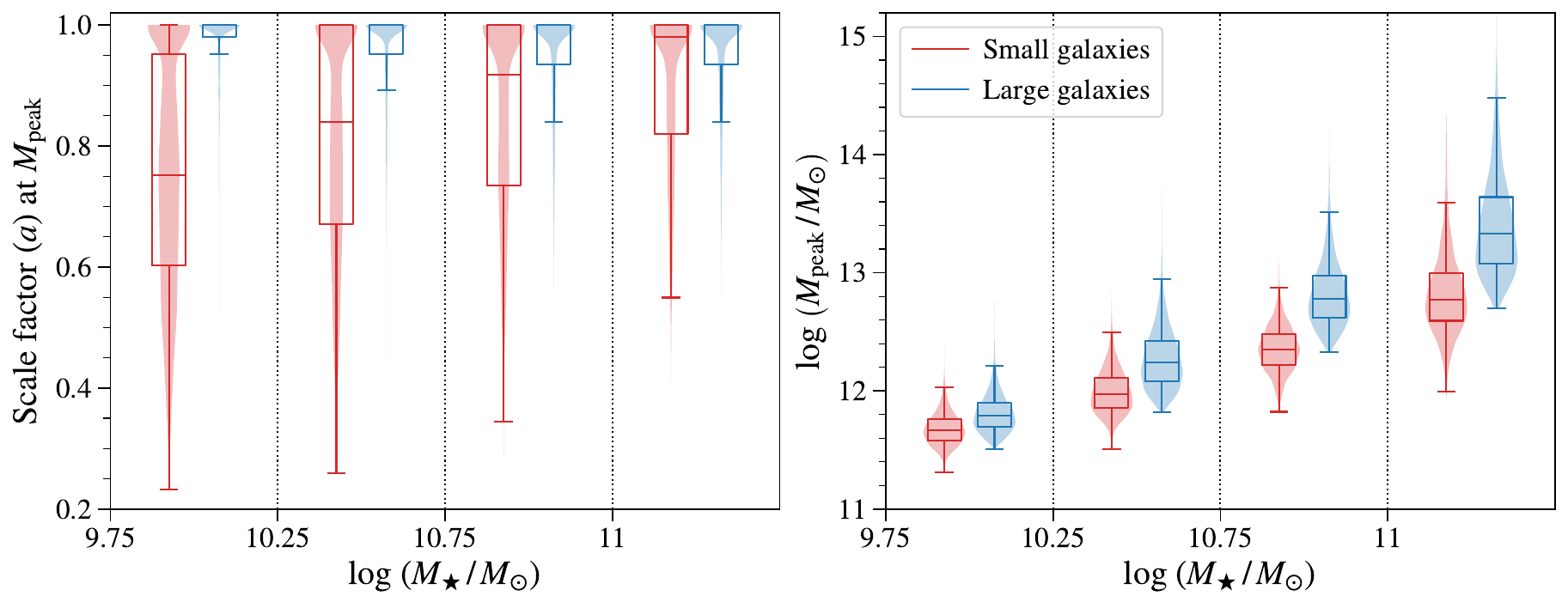}
\caption{Distributions of \ampeak{} (\textit{left}) and \mpeak{} (\textit{right}) of small (\textit{red}) and large (\textit{blue}) galaxies in bins of stellar mass, \logmstar, of $[9.75, 10.25)$, $[10.25, 10.75)$, $[10.75, 11)$, and $\geq 11$. The distribution (of \ampeak{} or \mpeak{}) is shown vertically as a violin plot, with a box plot overlaid to show the median, inner quartiles, and 1.5 times the interquartile range (IQR). The difference in \ampeak{} distributions between small and large galaxies contribute to stronger relative halo assembly bias, while the difference in \mpeak{} distributions contribute to stronger relative halo bias. \label{fig:ampeak_mpeak}}
\end{figure*}

\begin{figure}[tb!]
\centering
\includegraphics[width=\linewidth]{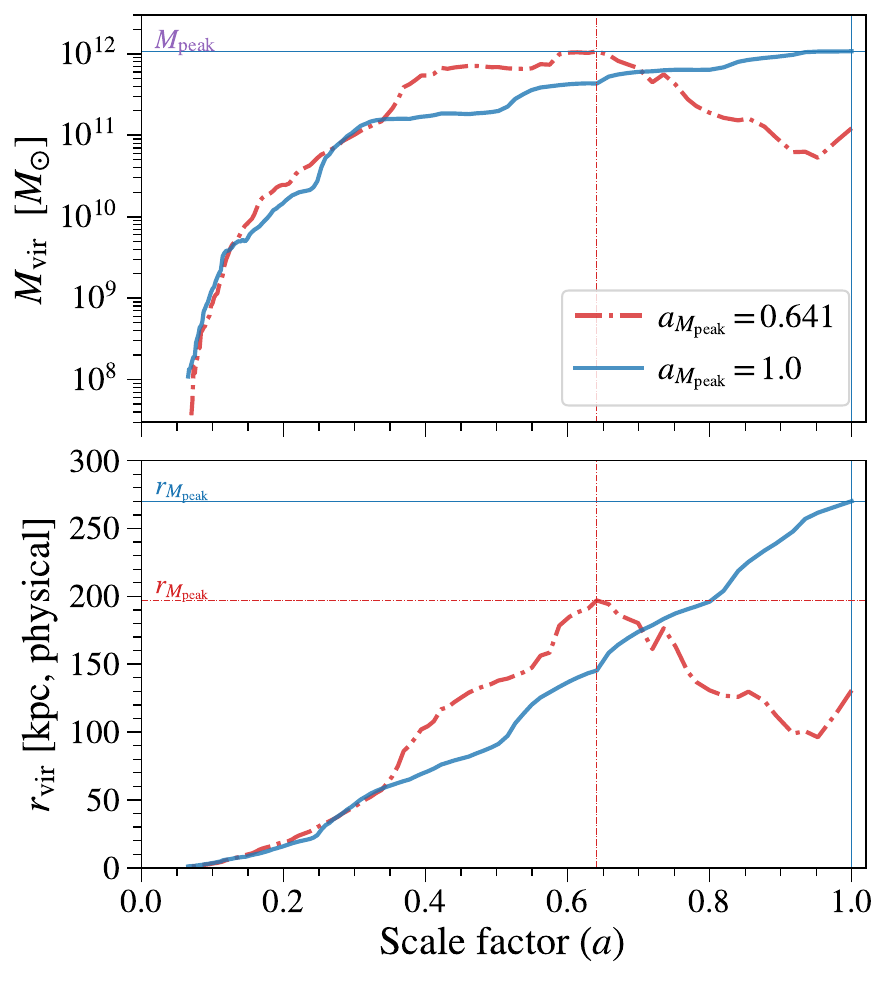}
\caption{The time evolution (time shown as scale factor, $a$) of halo mass (\mvir{}; \textit{upper}) and virial radius (\rvir{}; \textit{lower}) for two halos with the same \mpeak{} but different \ampeak{} values: $\ampeak{}= 0.641$ (\textit{red dash-dotted curves}) and 1 (\textit{blue solid curves}). 
The \rvir{} values shown are converted to physical distances.
The thin horizontal lines indicate the corresponding \mpeak{} and \rmpeak{} values, while the vertical lines indicate the corresponding \ampeak{} values. 
Note that even though the two halos have the same \mpeak{} value, the halo that reaches \mpeak{} at an earlier \ampeak{} results in a smaller \rmpeak{}.} 
\label{fig:twohalos}
\end{figure}

Relative halo assembly bias between the large and small galaxies play a major role in their respective clustering signals, especially at low masses. To see the direct role of assembly bias we look at the \ampeak{} distributions of large and small galaxies as a function of stellar mass. The left panel of \autoref{fig:ampeak_mpeak} shows the \ampeak{} distribution of large and small galaxies in four disjoint stellar mass bins: $9.75 \leq \logmstar \leq 10.25$, $10.25 \leq \logmstar \leq 10.75$, $10.75 \leq \logmstar \leq 11$, $\logmstar \ge 11$. In each stellar mass bin, the distributions of \ampeak{} are plotted in the violin plot style with a box plot overlaid to show the median, inner quartiles, and 1.5 times the interquartile range (IQR). 
We can clearly observe that small galaxies reside in halos with lower (i.e., earlier) \ampeak{} values compared to large galaxies. 

The earlier \ampeak distribution of the small galaxies imply those small galaxies tend to be assigned to halos that form earlier and have been stripped, either because they have become a subhalo of a larger system, or because they had passed near a larger system and become tidally disturbed. 
At a fixed halo mass, the halos that form earlier or having earlier \ampeak are more clustered; this is known as the halo assembly bias effect \citep{Wechsler:2006,2020MNRAS.493.4763M}. As such, the halo assembly bias leads to the higher clustering signal in small galaxies that we observed in the upper panel of \autoref{fig:clustering}. 

As the \ampeak distributions of the small and large galaxies become more similar as the stellar mass increases, the relative halo assembly bias of small and large galaxies also decreases. However, even at the highest mass bin ($\logmstar \ge 11$), there is still a significant difference in the \ampeak distributions of the small and large galaxies, so the halo assembly bias should still play a role even in the highest mass bin. We will discuss this point in more detail in \autoref{sec:halobias}. 

Given the simple galaxy size model (\autoref{sec:galmodel}), one might wonder why the halo assembly bias would play a role at all. After all, the size model does not involve any formation time information. Below we show that by using \mpeak{} for the subhalo abundance matching technique (which generates stellar masses), the formation time information implicitly enters the galaxy size model. 

Imagine two halos with the same \mpeak{} but different \ampeak{}, as illustrated in \autoref{fig:twohalos}. 
The upper panel plots the virial mass, while the lower panel plots the corresponding virial radius, both as functions of the scale factor. The solid red lines represent the halo that reaches its peak mass at an earlier scale factor of $\ampeak{} = 0.641$, denoted by the vertical dashed red line. The solid blue lines correspond to the halo that reaches its \mpeak{} at $\ampeak{} = 1$, indicated by the vertical dot-dashed blue lines. The blue dot-dashed horizontal lines mark the \mpeak{} and \rmpeak{} of the blue halo in the left and right panels, respectively, while the red dashed lines indicate the same for the red halo. 
 
Since \rmpeak{} increases as \ampeak{} increases, the halo with an earlier \ampeak{} will have a smaller \rmpeak{} than the later-forming halo. Recall that \autoref{eq:mvir} (the \mvir{}--\rvir{} relation) has a time dependence, which in turn introduces the time dependence into \autoref{eq:rhalf} . Consequently, \autoref{eq:rhalf} assigns smaller galaxies to earlier-forming halos at a fixed \mpeak{} value.  In other words, the combination of using \mpeak{} in abundance matching and using \rmpeak{} in the size model results in the dependence of galaxy sizes on halo formation time, even at a fixed stellar/halo mass, making halo assembly bias directly contribute to the clustering signal. 

\subsection{Halo Bias Effect}
\label{sec:halobias}

As discussed in \autoref{sec:assemblybias}, the relative halo assembly bias is a major factor that causes small galaxies (with earlier \ampeak{} values) to cluster more than large galaxies. Given the distributions shown in the left panel of \autoref{fig:ampeak_mpeak}, the halo assembly bias effect should exist for all four stellar mass bins in our analysis.  However, \autoref{fig:clustering} shows that, at $z=0$, large and small galaxies cluster similarly at high masses. Here we will demonstrate that the reason behind this phenomenon is the counter effect of halo bias. 

To understand the strength of the relative halo bias signal we look at the \mpeak{} distributions of large and small galaxies as a function of \mstar{}. These distributions are shown in the right panel of \autoref{fig:ampeak_mpeak}, for four disjoint \mstar{} bins: $9.75 \leq \logmstar \leq 10.25$, $10.25 \leq \logmstar \leq 10.75$, $10.75 \leq \logmstar \leq 11$, and $\logmstar \ge 11$. In each stellar mass bin, the distributions of \mpeak{} are plotted in the violin plot style with a box plot overlaid to show the median, inner quartiles, and 1.5\,IQR.

For the lower stellar mass bins in the right panel, the \mpeak{} distribution of small and large galaxies has significant overlap, meaning that for low-mass galaxies, large and small galaxies occupy halos of similar halo masses. In other words, at low stellar masses, the relative halo bias between large and small galaxies is minimal. 
On the other hand, for the higher stellar mass bins, the overlap between the \mpeak{} distributions of small and large galaxies diminishes.  As stellar mass increases, larger galaxies start to occupy more massive halos compared to small galaxies, resulting in the widening of the difference in \mpeak{}. This behavior is due to the shallower slope at the high-mass end of the SHMR, which can already be seen in \autoref{fig:SMHR} (see also the discussion in \autoref{sec:galclustering}). The right panel of \autoref{fig:ampeak_mpeak} demonstrate this behavior explicitly.

The further apart the \mpeak{} distributions between the small and large galaxies, the stronger the relative halo bias effect there is between the two subsamples. As halos of higher masses cluster more significantly than halos of lower masses (this is the definition of halo bias; see \citealt{1999MNRAS.308..119S}), large galaxies (occupying higher mass halos at a fixed stellar mass) will be more clustered due to halo bias. 
However, as demonstrated in the right panel of \autoref{fig:ampeak_mpeak}, this halo bias effect is only important for high-mass galaxies.

\subsection{Combination of both effects}
\label{sec:botheffects}

To summarize, we found that the \ampeak{} distributions of the small and large galaxies differ significantly at all stellar masses, with small galaxies having earlier \ampeak{}, resulting in halo assembly bias boosting the clustering signals of small galaxies at all masses.  We also found that the \mpeak{} distributions of small and large galaxies differ significantly only at high masses, with larger galaxies having higher \mpeak{}, resulting in halo bias boosting the clustering signals of larger galaxies but only at high masses. 
Consequently, at lower stellar masses, small galaxies are more clustered overall. As the stellar mass increases, the halo bias starts to increase the clustering of the large galaxies, and in the end, causes the large and small galaxies to cluster similarly at the highest stellar mass thresholds. 

Note that in the upper ($z=0$) panels of \autoref{fig:clustering}, the clustering of small galaxies (red lines) remains at a similar strength for the different stellar mass threshold increases. Whereas the clustering of large galaxies (blue lines) is lower in the left panel and increases as the stellar mass threshold increases. In other words, the small galaxies cluster similarly across all stellar mass thresholds, while the large galaxies cluster more as the stellar mass increases. This is consistent with our finding above because, for small galaxies, as stellar mass increases, the slight decrease in halo assembly bias is offset by the increase in halo bias, leading them to cluster similarly at all stellar masses. For large galaxies, however, the halo assembly bias remains similar at different stellar mass bins but the halo bias increases with stellar mass, so their overall clustering signal increases with stellar mass. 

For another exercise that demonstrates the same effect, please see \autoref{sec:appendix}, where we manually remove the relative halo assembly bias between the small and large subsamples, leaving only the halo bias effect. In this case, we found that large galaxies are always more clustered than small galaxies, and the clustering gap (due to the remaining halo bias) increases with stellar mass as we expect.

\subsection{Choice of \texorpdfstring{\rmpeak{} vs.\@ \rvir{}}{Rmpeak vs. Rvir}}
\label{sec:rvir}

One of our main findings from \autoref{sec:assemblybias} is that, due to the time dependence in the \mvir{}--\rvir{} relation, the \rmpeak{} size model implicitly encodes the halo formation time into the modeled sizes, which in turn results in the clustering difference between the small and large galaxies. One natural follow-up question to ask is whether this halo assembly bias effect would disappear if we always use \rvir{} in the size model instead (i.e., galaxy size is proportional to present-day \rvir{} regardless of the value of \ampeak{}). It turns out that the halo assembly bias exists in both the \rvir{} and \rmpeak{} models; it simply arises from two related but different mechanisms from the halo assembly history. 

Consider again, two halos with the same \mpeak{} but different \ampeak{} values, as shown in \autoref{fig:twohalos}. In the \rmpeak{} size model (\autoref{eq:rhalf}), the halo that reaches its \ampeak{} earlier will be assigned a smaller galaxy compared to the halo that reaches its \ampeak{} later, resulting in assembly bias in the smaller galaxy samples at a fixed \mpeak. Now, if we use \rvir{} instead of \rmpeak{} in \autoref{eq:rhalf}, halos with similar \rvir{} at $a = 1$ would be assigned similar-sized galaxies. However, if the two halos end up with different \rvir{} values, the halo with the smaller \rvir{} will be assigned a smaller galaxy. This difference in \rvir{} indicates that at least one halo has undergone tidal stripping more than the other after reaching its peak mass. Consequently, tidal stripping introduces a scale factor dependence in the \rvir{} case, contributing to the halo assembly bias effect for smaller galaxies.

Somewhat surprisingly, these two size models (using \rmpeak{} vs.\@ \rvir{}) result in very similar clustering signals. In \autoref{fig:clustering}, the dashed lines show the clustering signal for the \rvir{} model. At $z=0$ (upper panels of \autoref{fig:clustering}), the clustering gap between small and large galaxies is slightly smaller for the \rvir{} model, but the difference is barely discernible.
The main reason for the similarity between the two models' clustering signals is that the \ampeak{} and \mpeak{} distributions for each galaxy subsample are very similar between these two models. In other words, when we remake \autoref{fig:ampeak_mpeak} but with the \rvir{} model instead, the resulting figure looks very similar. We include this figure as \autoref{fig:ampeak_mpeak_rvir} in \autoref{sec:appendix} for readers who wish to compare them. 
In \autoref{sec:highz-rmpeak-rvir}, we further investigate if higher-redshift data may help distinguish the \rmpeak{} and \rvir{} models.

This result is consistent with the finding of \citet{2101.05280}. They found that the clustering signals alone, especially if evaluated only at a single time, may not be sufficient to distinguish which dark matter halo properties are most tightly connected to an observable galaxy property (galaxy size in our case). In fact, many secondary halo properties, including various characteristics of the halo assembly history, can result in comparable halo assembly bias signals (hence, sometimes referred to as secondary halo bias; cf.\@ \citealt{2018MNRAS.474.5143M}). As long as the galaxy size model introduces a dependence on a secondary halo property that results in a clustering gap mimicking those observed in red data sets, the model would remain a plausible choice under the galaxy clustering test. 

\section{Discussion}
\label{sec:discussion}

\subsection{Comparison with Alternative Models}
\label{sec:compare-models}
In \cite{2019MNRAS.488.4801J}, they proposed alternative galaxy size models that explicitly include the dependence on halo assembly history. Specifically, their model takes the form of 
\begin{align}
\rhalf = 0.02(c/10)^{-0.7}\rvir{}
\end{align}
where $c$ is the halo's concentration parameter (the ratio of the virial radius to the scale radius). Because the concentration parameter is known to be highly correlated with halo formation time \citep{Wechsler:2002,Mao:2015}, the \cite{2019MNRAS.488.4801J} model will produce different galaxy sizes for halos that have the same $\rvir{}$.  
Similarly, \citet{2101.05280} proposed a galaxy size model that incorporates halo growth rate. In their model, the size of a galaxy is correlated with both $\rvir{}$ and the average mass accretion rate over the past dynamical time $\dot M_h$. Therefore,  halos that have the same $\rvir{}$ can also host different galaxy sizes in this model.

In contrast, for the \rvir{} (or \rmpeak{}) model discussed in this work, halos that have the same $\rvir{}$ (or \rmpeak{}) will have the same galaxy sizes. However, when comparing with observations, we mostly care about the galaxy sizes at a fixed stellar mass, not at a fixed halo mass, because the latter is not observable. The main finding of this work is that, even if a simple (linear) galaxy size--halo radius relation does not depend on the halo assembly history, the resulting galaxy size--\textit{stellar} mass relation would still depend on halo assembly history. 

In fact, the observation of small galaxies being more clustered than large galaxies at low stellar masses directly implies that there must exist some dependence, either explicit or implicit, on the halo assembly history. In a fixed stellar mass bin, the small galaxies must occupy halos that either are more massive (more clustered due to halo bias) or form early (more clustered due to halo assembly bias). Since it would be rather unphysical to have small galaxies occupy halos with larger halo radii, we arrive at the conclusion that small galaxies have to occupy halos that form early. 

This conclusion highlights both a challenge and an opportunity. The main challenge, as discussed in \autoref{sec:rvir}, is the difficulty in identifying the specific halo properties that control galaxy sizes. The simple \citetalias{Kravtsov:2013} model and the models that have explicit dependence on halo assembly history (e.g., in \citealt{2019MNRAS.488.4801J} and \citealt{2101.05280}) can all reproduce reasonable size--mass relation and clustering signals. 
On the flip side, as we consider future improvement to the galaxy size model, a simple model (without explicit dependence on the halo assembly history) may work as well as ones that have. For example, one can explore the non-linear dependence on the halo radius without including a new dependence on the halo concentration or formation time.

\subsection{Size--mass Relations at Higher Redshifts}
\label{sec:highzpred}

\begin{figure}[tb!]
\centering
\includegraphics[width=\linewidth]{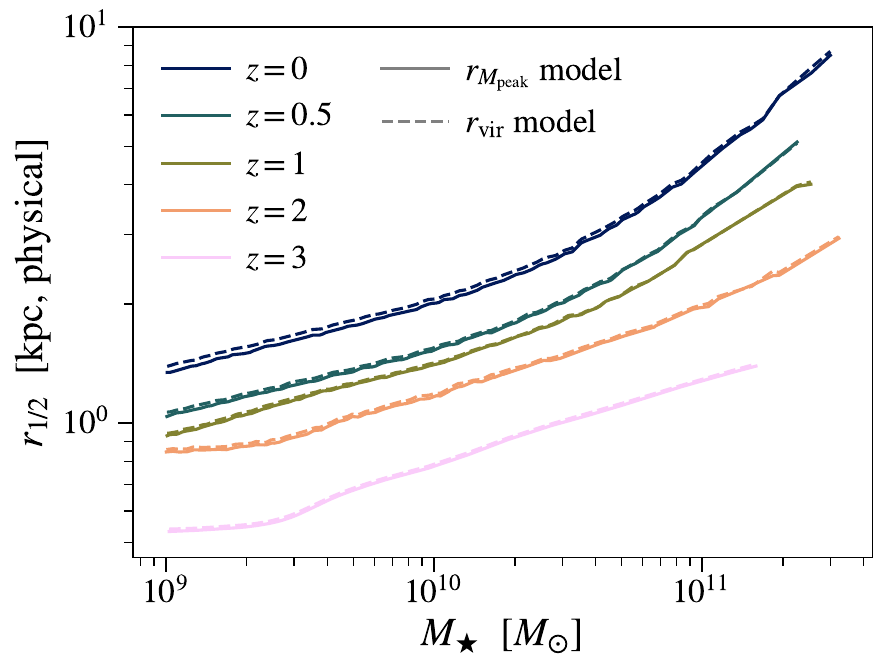}
\caption{Model-predicted relation between physical half light radius and stellar mass (\rhalf{}--\mstar{} relation) at multiple redshifts (from top to bottom, $z=0$, 0.5, 1, 2, and 3; also shown as different shades of colors from dark blue to light pink).  The corresponding dashed lines denote the same redshifts but using the \rvir{} model. Note that the proportional constant in the size model is held fixed across redshifts for both model choices.}
\label{fig:highz}
\end{figure}

Following up on our discussion in \autoref{sec:rvir}, one possible way to further distinguish different galaxy size models is to compare their high-redshift (early-time) predictions. If the assembly bias effect has a time dependence that differs in different models, we should be able to see different high-redshift behaviors, which can then be compared with observed data. In this section, we compare the high-redshift predictions from the \rmpeak{} and \rvir{} galaxy size models. 

To produce high-redshift predictions, we use the same modeling recipe detailed in \autoref{sec:model}. We use the 
halo catalogs at $z = 0.5$, 1, 2, and 3, and also change the stellar mass function used during the abundance matching step (for generating stellar masses) to the ones at the corresponding redshifts. The high-redshift stellar mass functions used are from \citet[for $z = 0.5$]{Moustakas:2013} and \citet[for $z = 1$, 2, and 3]{2011MNRAS.413.2845M}; these stellar mass functions are the same ones used in \citet{2013ApJ...770...57B}. 
We follow the same procedure as \autoref{sec:galmodel} to generate galaxy sizes (with either \rmpeak{} or \rvir{}), fixing the proportional constant at 0.01 regardless of the redshift values.   

\autoref{fig:highz} presents the predicted median \rhalf{}--\mstar{} relations at all five redshifts, $z \in [0, 0.5, 1, 2, 3]$, from both the \rmpeak{} and \rvir{} size models. The \rhalf{} values are reported in physical sizes. First, we see that at a fixed stellar mass, the median galaxy size increases with time significantly. The size evolution with time predicted by the simple size model appears to be greater than one would expect \citep{Huang:2017,2024ApJ...972..134M}, suggesting that the proportional constant in the galaxy size model likely need to evolve (increase) with time as well. 
In fact, the models proposed by \cite{2019MNRAS.488.4801J},  indeed include a proportionality constant that decreases with time. They found the scaling $(1+z)^{0.5}$ fits the data the best.


\subsection{Can High-redshift Data Distinguish \texorpdfstring{\rmpeak{} vs.\@ \rvir{}}{Rmpeak vs. Rvir} Models?}
\label{sec:highz-rmpeak-rvir}

However, regardless of the choice of proportionality constant in the size model, the other important result shown in \autoref{fig:highz} is that the \rmpeak{} and \rvir{} models produce virtually identical \rhalf{}--\mstar{} relations at all redshifts considered. In fact, the very slight difference between the \rhalf{}--\mstar{} relations produced by the two models decreases with redshift.   Recall that the galaxy size is controlled by the value of \ampeak{} in the \rmpeak{} model, and by the amount of stripping in the \rvir{} model. The fact that these two models produce similar \rhalf{}--\mstar{} relations at different redshifts suggests that \ampeak{} and the amount of stripping are highly (anti-)correlated, and the correlation is preserved at higher redshifts as well. 

We further investigate the predicted clustering signals at higher redshifts. In the bottom panel of \autoref{fig:clustering}, the clustering signals of large and small galaxies at $z = 3$ for both model choices of (\rmpeak{} and \rvir{}) are displayed. First, we note that, for every stellar mass threshold, large galaxies cluster slightly more than small galaxies. This behavior is consistent with our explanation in \autoref{sec:results}, where we argue that in the absence of assembly bias, large galaxies should be more clustered due to the halo bias effect that comes from the small halo mass difference. Because the halo assembly bias accumulates as the halos assemble with time, the halo assembly bias effect is much less prominent at $z = 3$ compared to that at $z=0$, resulting in the clustering signals we see in \autoref{fig:clustering}. 

Unfortunately, the \rmpeak{} and \rvir{} size models still produce identical clustering signals at $z=3$ (and at all other redshifts that we have analyzed). In fact, due to the weaker assembly bias effect at $z=3$, the tiny difference between the two models (solid and dashed lines) that we observe at $z=0$ for the lower stellar mass thresholds completely disappears at $z=3$. 
This result again echoes \citet{2101.05280}, showing that it can be quite challenging to identify a particular halo property that controls galaxy sizes, even with galaxy clustering constraints from multiple redshifts. 

Note that when we discuss the high-redshift predictions here, we did not consider galaxies of different morphologies at all, as this work focuses on the theoretical origins of the clustering signals only. In reality, the size--mass relation for late- and early-type galaxies can be different (cf. Figure~1 of \citetalias{Kravtsov:2013}). As galaxy morphology is also connected to the assembly history of the system, it can play a role in the clustering signals as well. This investigation can be an avenue for future work.

\section{Summary}
\label{sec:summary}

This work provides a quantitative explanation for size-split galaxy clustering signals seen in \citetalias{Hearin:2019}, where the galaxy sizes are assigned using the \citetalias{Kravtsov:2013} model.  In particular, we answer the question of why the model can produce a galaxy clustering gap between large and small galaxies without explicitly involving halo formation time in the size model. 
We showed that the clustering pattern originates from the counter effects between halo assembly bias and halo bias as functions of stellar mass. 
At lower stellar masses, we showed that small and large must occupy halos of different assembly histories, otherwise they cannot result in the observed clustering difference.
Because the effect of halo assembly bias can enter the model even when the model does not explicitly depend on assembly history, our result highlights the challenge of constraining the galaxy--halo size relation model from galaxy clustering alone. 
Our key findings can be summarized as follows:

\textit{Effects of Halo Assembly Bias and Halo Bias}--- Halo assembly bias effect makes small galaxies more clustered as they have earlier \ampeak{} values, due to the size model's choice of \rmpeak{} (\autoref{sec:assemblybias}). At higher stellar masses, the shallower SMHR makes galaxies at a fixed stellar mass occupy a wider range of halo mass, resulting in a more significant relative halo bias between the small and large galaxies (\autoref{sec:halobias}). The combined effect explains why small galaxies cluster more at lower stellar masses, while at high masses, the clustering of large and small galaxies converges as halo bias cancels out the assembly bias effect (\autoref{sec:botheffects}). 

\textit{Prediction at Higher Redshifts}--- When the proportionality constant used in the size model is held fixed with time, the \rhalf{}--\mstar{} relation decreases with increasing redshift (i.e., size increases with time at a fixed mass). Compared with the little evolution in the observed \rhalf{}--\mstar{} relation, suggesting the proportionality constant in the size model needs to evolve with time. The galaxy clustering signals at higher redshifts show a weaker halo assembly bias effect, consistent with theoretical expectation. 

\textit{The choice of \rmpeak{} vs.\ \rvir{}}--- Using \rmpeak{} and \rvir{} in the \citetalias{Kravtsov:2013} galaxy size model produce almost identical results, including the size-split galaxy clustering signals and \rhalf{}--\mstar{} relations at multiple redshifts. The tidal stripping introduces the dependence on halo assembly history into the \rvir{} model.
The similarity between the two models is due to the strong (anti-) correlation of the \ampeak{} value and the amount of stripping after the halo reaches its peak mass. 

This result demonstrates that, if a galaxy size model that reproduces the observed clustering pattern, where small galaxies cluster more at low stellar masses (as shown in \autoref{fig:clustering}), then the model must include some degree of the halo assembly bias effect. Otherwise, large galaxies would exhibit stronger clustering signals due to the halo bias effect. In other words, at a fixed stellar mass, the small galaxies must live in halos that form earlier.

With a greater theoretical understanding of how the galaxy size model produces the galaxy clustering signals, we hope to develop new observational constraints and improve the model to produce better predictions in future work. 

\acknowledgments{

The authors thank Peter Behroozi, Andrew Hearin, and Andrey Kravtsov for their helpful comments. 

This work made use of the VSMDPL simulation. 
The authors gratefully acknowledge the Gauss Centre for Supercomputing e.V. (www.gauss-centre.eu) and the Partnership for Advanced Supercomputing in Europe (PRACE, www.prace-ri.eu) for funding the MultiDark simulation project by providing computing time on the GCS Supercomputer SuperMUC at Leibniz Supercomputing Centre (LRZ, www.lrz.de).
The MultiDark Planck simulation suite has been performed by Gustavo Yepes at the SuperMUC supercomputer at LRZ using time granted by PRACE and project time granted by LRZ. Processed data of these simulations were created by P. Behroozi, S. Gottlöber, A. Klypin, N. Libeskind, F. Prada, V. Turchaninov and G. Yepes.

The support and resources from the Center for High Performance Computing at the University of Utah are gratefully acknowledged.
This research has made use of NASA's Astrophysics Data System.
This work made use of the following publicly available Python packages: 
Numpy \citep{numpy},
SciPy \citep{scipy},
Matplotlib \citep{matplotlib},
IPython \citep{ipython},
Jupyter \citep{jupyter},
pandas \citep{2022zndo...3509134T},
Halotools \citep{1606.04106},
AbundanceMatching (\citealt{2022ascl.soft12016M}; \https{github.com/yymao/abundancematching}),
easyquery (\https{github.com/yymao/easyquery}), and
adstex (\https{github.com/yymao/adstex}).

}

\bibliography{references}

\appendix

\section{Additional Analyses}
\label{sec:appendix}

\begin{figure*}[tb!]
\centering
\includegraphics[width=\linewidth]{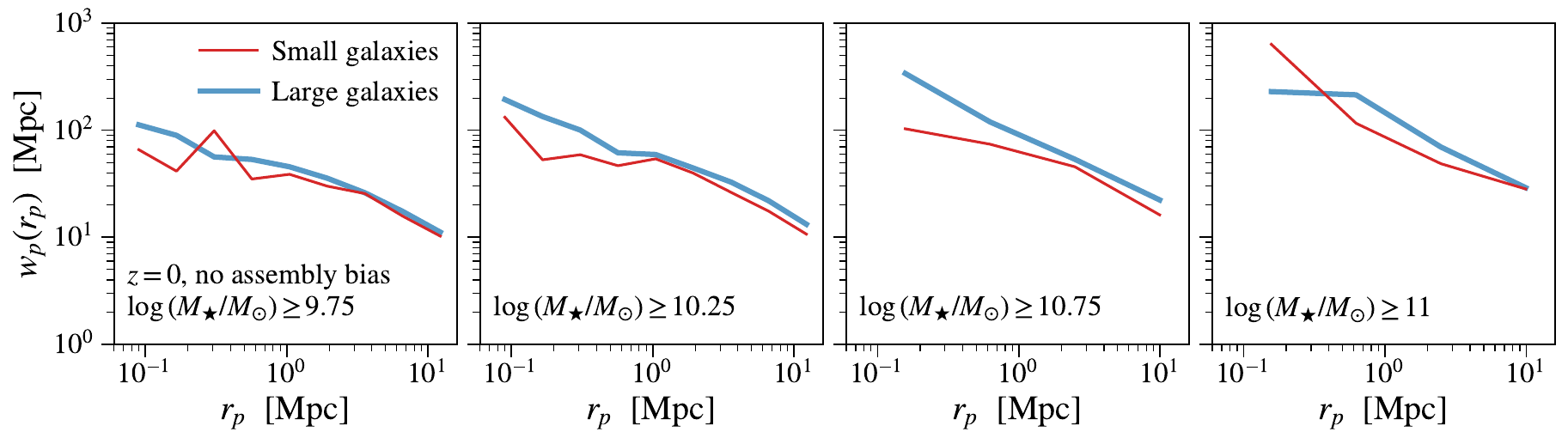}
\caption{Projected two-point (galaxy--galaxy) correlation function, $w_p(r_p)$, for galaxy subsamples of which the halo assembly bias is manually removed by including only halos with $\ampeak{}=1$ (see \autoref{sec:appendix} for the procedure).
The four columns show $w_p(r_p)$ of four galaxy samples corresponding to different stellar mass thresholds (\textit{from left to right}): $\logmstar \ge 9.75$, 10.25, 10.75, and 11.
The color and thickness of the curves denote the two size-split subsamples: small (\textit{red thin curves}) and large (above median; \textit{blue thick curves}) galaxies. The galaxy sizes are generated by the \rmpeak{} model.
\label{fig:matchedclustering}
}
\end{figure*}

\begin{figure*}[tb!]
\centering
\includegraphics[width=\linewidth]{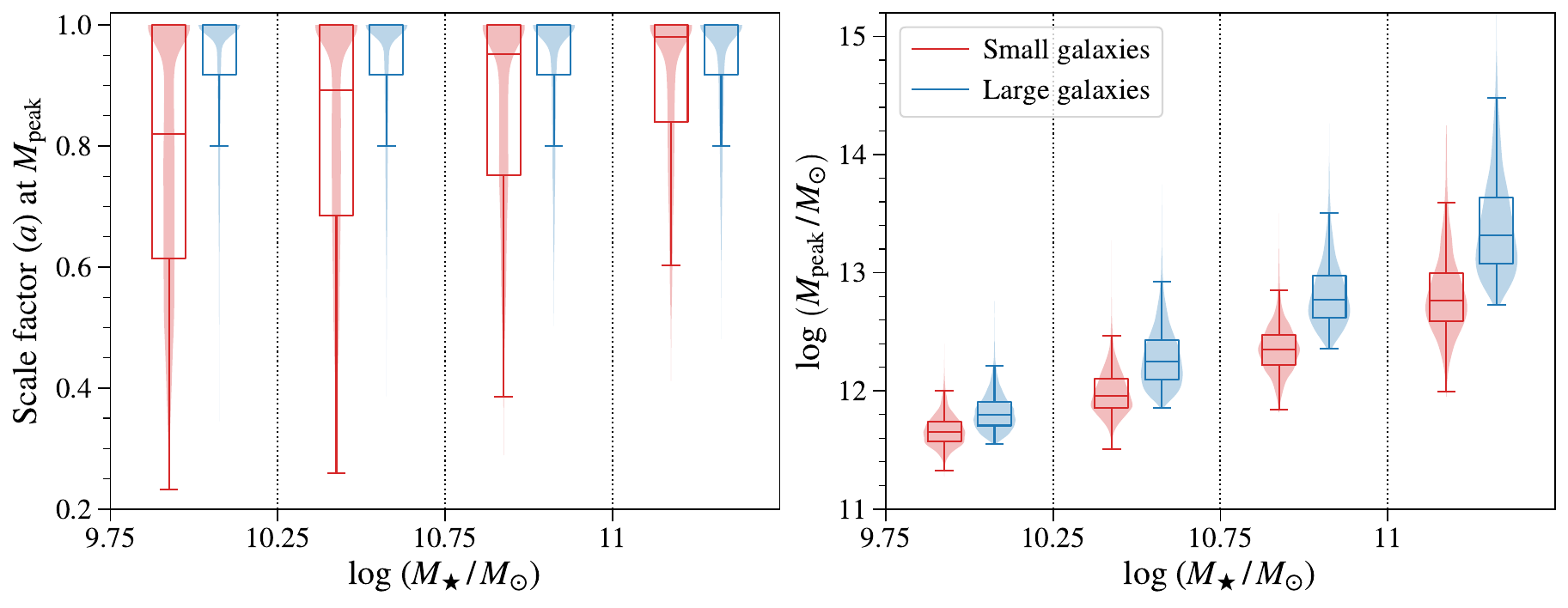}
\caption{Same as \autoref{fig:ampeak_mpeak} but using the \rvir{} model to generate galaxy sizes.
\label{fig:ampeak_mpeak_rvir}
}
\end{figure*}

In this appendix, we include two additional analyses for interested readers. First, following the discussion in \autoref{sec:botheffects}, we provide further evidence to support our conclusion that assembly bias has a major role in the clustering of large and small low-mass galaxies. We expect that, if one manually removes the halo assembly bias, the large galaxies should cluster slightly more in all four stellar mass samples, due to the halo bias effect.
To remove the halo assembly bias caused by \ampeak{}, we take a subsample of large and small galaxies that have $\ampeak{} = 1$. This effectively removes any $\ampeak{}$ assembly bias since every halo in the sub-sample reaches its \mpeak{} at the same time. We then weigh the \mpeak{} distributions for small galaxies and large galaxies to match their respective original \mpeak{} distributions to preserve any potential halo bias signals. This procedure produces the clustering signals seen in \autoref{fig:matchedclustering}, where now at all stellar mass thresholds, large and small galaxies cluster very similarly. Indeed, the large galaxies actually cluster slightly more as expected, especially at higher stellar masses where the halo bias effect is the strongest. This result provides further evidence that, at lower stellar masses, halo assembly bias is the main cause of small galaxies clustering more than large galaxies.

Second, we show the \ampeak{} and  \mpeak{} distributions for the \rvir{} model, as discussed in \autoref{sec:rvir}.  These distributions are shown in \autoref{fig:ampeak_mpeak_rvir}, which is identical to \autoref{fig:ampeak_mpeak} but using \rvir{} instead of \rmpeak{} in the size model. The choice of \rvir{} introduces assembly bias through tidal stripping, the more a halo has been stripped since it reaches peak mass, the more likely it has a smaller galaxy size assigned. We found the \ampeak{} and \mpeak{} distributions in \autoref{fig:ampeak_mpeak_rvir} (\rvir{} model) and \autoref{fig:ampeak_mpeak} (\rmpeak{} model) are very similar, indicating the strong correlation between the mass loss and \ampeak. This correlation is expected as more heavily stripped halos are likely to have an earlier \ampeak \citep{2018MNRAS.481.4038L}. 
Because the \rmpeak{} and \rvir{} models produce such similar \ampeak{} and  \mpeak{} distributions, the two models also produce almost identical clustering signals, as already discussed in \autoref{sec:rvir}. 

\bigskip

\end{document}